\documentclass[a4paper]{article}

\usepackage{INTERSPEECH2019}
\usepackage[acronym]{glossaries}
\usepackage{siunitx}
\usepackage{multirow}
\usepackage{multicol}
\usepackage{makecell}
\usepackage{subcaption}
\usepackage[table]{xcolor}
\usepackage[tracking=true]{microtype}

\makeatletter

\renewcommand{\section}{\@startsection
  {section}%
  {1}%
  {}%
  {0.9\baselineskip}%
  {0.2\baselineskip}%
  {}}%

\renewcommand{\subsection}{\@startsection
  {subsection}%
  {2}%
  {}%
  {-0.1\baselineskip}%
  {0.1\baselineskip}%
  {}}%

\renewcommand{\subsubsection}{\@startsection
  {subsubsection}%
  {3}%
  {}%
  {0.5\baselineskip}%
  {0.1\baselineskip}%
  {}}%

\def\thebibliography#1{\begin{center}
{\bf \large References\vspace{-.5em}\vspace{2pt}}
\end{center}\eightpt\list
 {[\arabic{enumi}]}{\settowidth\labelwidth{[#1]}\leftmargin\labelwidth
 \advance\leftmargin\labelsep
 \usecounter{enumi}}
 \def\newblock{\hskip .11em plus .33em minus .07em}
 \sloppy\clubpenalty4000\widowpenalty4000
 \sfcode`\.=1000\relax}

\g@addto@macro\normalsize{%
  \setlength\abovedisplayskip{4pt plus 2pt minus 2pt}
  \setlength\belowdisplayskip{4pt plus 2pt minus 2pt}
  \setlength\abovedisplayshortskip{4pt plus 2pt minus 2pt}
  \setlength\belowdisplayshortskip{4pt plus 2pt minus 2pt}
}

\makeatother

\newacronym{AM}{AM}{acoustic model}
\newacronym[plural=ANNs, firstplural=artificial neural networks (ANNs)]{ANN}{ANN}{artificial neural network}
\newacronym{ASR}{ASR}{automatic speech recognition}
\newacronym{BAN}{BAN}{blind analytic normalization}
\newacronym{BSS}{BSS}{\#TBD !!!}
\newacronym[plural=BLSTMs, firstplural=bidirectional long short-term memories (BLSTMs)]{BLSTM}{BLSTM}{bidirectional long short-term memory}
\newacronym{CE}{CE}{cross entropy}
\newacronym{DAS}{DAS}{delay and sum}
\newacronym{DFT}{DFT}{discrete Fourier transform}
\newacronym{DM}{DM}{direct masking}
\newacronym{DPCL}{DPCL}{deep clustering}
\newacronym{FS}{FS}{filter and sum}
\newacronym[plural=GMMs, firstplural=Gaussian mixture models (GMMs)]{GMM}{GMM}{Gaussian mixture model}
\newacronym[plural=DNNs, firstplural=deep neural networks (DNNs)]{DNN}{DNN}{deep neural network}
\newacronym{GEV}{GEV}{generalized eigenvalue}
\newacronym{PESQ}{PESQ}{perceptual evaluation of speech quality}
\newacronym{PIT}{PIT}{permutation invariant training}
\newacronym{PW}{PW}{parametric Wiener filter}
\newacronym[plural=HMMs, firstplural=hidden Markov models (HMMs)]{HMM}{HMM}{hidden Markov model}
\newacronym[plural=LSTMs, firstplural=long short-term memories (LSTMs)]{LSTM}{LSTM}{long short-term memory}
\newacronym{MVDR}{MVDR}{minimum variance distortionless response}
\newacronym[plural=RNNs, firstplural=recurrent neural networks (RNNs)]{RNN}{RNN}{recurrent neural network}
\newacronym{SGD}{SGD}{stochastic gradient descent}
\newacronym{SDR}{SDR}{signal to distortion ratio}
\newacronym{SNR}{SNR}{signal-to-noise ratio}
\newacronym{STFT}{STFT}{short-time Fourier transform}
\newacronym[plural=WERs, firstplural=word error rates (WERs)]{WER}{WER}{word error rate}
\newacronym{WSJ}{WSJ}{Wall Street Journal}

\newcommand{\wsjOrigDevNinetythreeFiveK}{4.6}
\newcommand{\wsjOrigEvalNinetytwoFiveK}{1.8}
\newcommand{\wsjOrigEvalNinetythreeFiveK}{4.0}
\newcommand{\wsjOrigDevNinetythreeTwentyK}{10.0}
\newcommand{\wsjOrigEvalNinetytwoTwentyK}{6.9}
\newcommand{\wsjOrigEvalNinetythreeTwentyK}{9.4}

\newcommand{\dpclGmmHmmEvalA}{30.8}

\newcommand{\EtEOneEvalA}{28.2}

\newcommand{\EtETwoEvalA}{25.4}

\newcommand{\dpclDnnHmmEvalA}{16.5} 
\newcommand{\dpclDnnHmmSameDomEvalA}{17.6} 
\newcommand{\dpclDnnHmmSameNondomEvalA}{22.2} 
\newcommand{\dpclDnnHmmDiffDomEvalA}{12.1} 
\newcommand{\dpclDnnHmmDiffNondomEvalA}{14.0} 
\newcommand{\dpclDnnHmmSameDomEvalASdr}{10.7} 
\newcommand{\dpclDnnHmmSameNondomEvalASdr}{8.0} 
\newcommand{\dpclDnnHmmDiffDomEvalASdr}{12.7} 
\newcommand{\dpclDnnHmmDiffNondomEvalASdr}{10.1} 

\title{Analysis of Deep Clustering as Preprocessing for\\ Automatic Speech Recognition of Sparsely Overlapping Speech}
\name{Tobias Menne$^{1,2}$, Ilya Sklyar$^1$, Ralf Schl\"{u}ter$^1$, Hermann Ney$^{1,2}$}
\address{
  $^1$Human Language Technology and Pattern Recognition, Computer Science Department, \\RWTH Aachen University, 52074 Aachen, Germany\\
  $^2$AppTek GmbH, 52062 Aachen, Germany
  }
\email{\{menne, schlueter, ney\}@cs.rwth-aachen.de\\sklyar@i6.informatik.rwth-aachen.de}

\begin{document}

\maketitle
\begin{abstract}
Significant performance degradation of \gls{ASR} systems is observed when the audio signal contains cross-talk. 
One of the recently proposed approaches to solve the problem of multi-speaker \gls{ASR} is the \gls{DPCL} approach.
Combining \gls{DPCL} with a state-of-the-art hybrid acoustic model, we obtain a \gls{WER} of \SI{\dpclDnnHmmEvalA}{\%} on the commonly used wsj0-2mix dataset, which is the best performance reported thus far to the best of our knowledge.
The wsj0-2mix dataset contains simulated cross-talk where the speech of multiple speakers overlaps for almost the entire utterance. 
In a more realistic \gls{ASR} scenario the audio signal contains significant portions of single-speaker speech and only part of the signal contains speech of multiple competing speakers.
This paper investigates obstacles of applying \gls{DPCL} as a preprocessing method for \gls{ASR} in such a scenario of sparsely overlapping speech.
To this end we present a data simulation approach, closely related to the wsj0-2mix dataset, generating sparsely overlapping speech datasets of arbitrary overlap ratio.
The analysis of applying \gls{DPCL} to sparsely overlapping speech is an important interim step between the fully overlapping datasets like \mbox{wsj0-2mix} and more realistic \gls{ASR} datasets, such as CHiME-5 or AMI.
\end{abstract}

\noindent\textbf{Index Terms}: deep clustering, ASR, speaker separation, multi-speaker ASR 

\section{Introduction}

The performance of \gls{ASR} systems on relatively clean, close-talking recordings has been improved drastically over the recent years.
This scenario can e.g. be found in telephony speech or readings of audio books.
On standard tasks for this scenario, as switchboard and librispeech \cite{theMic2017ConSpeRecSys, theCap2017ConvSpeRecSys}, typical \glspl{WER} are below \SI{10}{\%} .
Nevertheless, \gls{ASR} on noisy data remains challenging.
\gls{ASR} performance decreases especially when audio is recorded from a larger distance or when multiple speakers are talking simultaneously \cite{theFifChiSpeSepAndRecChaDatTasAndBas, theAmiMeeCorAPreAnn}.
There has thus been a growing interest in multi-speaker \gls{ASR}.
A special focus in this area lies on single-channel recordings \cite{perInvTraOfDeeModForSpeIndMulTalSpeSep, aPurEndToEndSysForMulSpeSpeRec, endToEndMonMulSpeAsrSysWitPre}.
This scenario is not only of interest if only one recording channel can be obtained, but also if multi-channel processing steps, like beamforming, can not separate two speakers because they are spatially too close to each other.

In \cite{aPurEndToEndSysForMulSpeSpeRec} a purely end-to-end system has been proposed which aims at directly recognizing multiple speakers with an attention based sequence-to-sequence model.
This systems employs no separate source separation stage.
Other recently proposed solutions use a separate source separation stage.
This preprocessing usually employs masks in the time-frequency domain to separate multiple sources from the mixture signal.
One method to obtain those masks, is to directly infer them using an \gls{ANN}.
This \gls{ANN} is trained by a permutation-free objective function \cite{deeCluDisEmbForSegAndSep, perInvTraOfDeeModForSpeIndMulTalSpeSep, mulTalSpeSepWitUttLevPerInvTraOfDeeRecNeuNet}. 
The method is often referred to as \gls{PIT}.
The integrated application of \gls{PIT} on the level of the \gls{ASR} cost function is presented in \cite{recMulTalSpeWitPerInvTra}.
A different approach to obtain the masks, is through the utilization of embedding vectors in the time-frequency domain.
This is done in the \gls{DPCL} and deep attractor network approaches \cite{deeCluDisEmbForSegAndSep, sinChaMulSpeSepUsiDeeClu, deeAttNetForSinMicSpeSep}.
The focus of the work presented here lies on the \gls{DPCL} approach. 
In \gls{DPCL} a \gls{ANN} is trained to map each time-frequency bin to an embedding vector.
Those embeddings are then used to allocate the time-frequency bins to different speakers through k-means clustering.
The objective function is designed such that embedding vectors of time-frequency bins belonging to the same speaker are close to each other.
Embeddings belonging to different speakers have a larger distance.

\gls{DPCL} has shown good potential when used in a preprocessing step for an \gls{ASR} system based on a \gls{GMM}-\gls{HMM} acoustic model \cite{sinChaMulSpeSepUsiDeeClu}.
Good performance was also achieved in combination with a sequence-to-sequence \gls{ASR} system \cite{endToEndMulSpeSpeRec}.
Results on the combination with a state-of-the-art hybrid \gls{DNN}-\gls{HMM} system have, to the best of our knowledge, not been published so far. 
Thus \gls{DPCL} tends to perform worse than the sequence-to-sequence and \gls{PIT} approaches in recent comparisons on wsj0-2mix \cite{aPurEndToEndSysForMulSpeSpeRec, endToEndMonMulSpeAsrSysWitPre}.
By combining \gls{DPCL} with a state of the art hybrid \gls{DNN}-\gls{HMM} we obtain a \gls{WER} of \SI{\dpclDnnHmmEvalA}{\%}, which to the best of our knowledge is the best performance reported on wsj0-2mix thus far.

Furthermore we present an in depth analysis of \gls{DPCL} as a preprocessing step for \gls{ASR} on a more realistic scenario of only sparsely overlapping speech. 
In this scenario only a small part of the signal contains multi-speaker segments whereas the majority of the signal contains single-speaker segments.
This scenario is much closer to the scenarios of meeting room or smart home recordings.
To this end we introduce a data simulation approach to obtain sparsely overlapping speech with a fixed but configurable overlap ratio based on the \gls{WSJ} data.
This data bridges the gap between \mbox{wsj0-2mix} and more realistic datasets as for example Chime-5 \cite{theFifChiSpeSepAndRecChaDatTasAndBas} and AMI \cite{theAmiMeeCorAPreAnn}.
The use of a simulated dataset offers the advantage of investigating various overlap ratios.
Furthermore it allows the utilization of oracle knowledge to analyze separate aspects of the pipeline and analyze their influence on \gls{ASR} performance in a controlled environment. 
Our analysis points out obstacles of applying \gls{DPCL} to sparsely overlapping data which are cloaked in the experiments on fully overlapping signals.
We also propose initial solutions.

The paper is organized as follows. 
An overview of the datasets used in this work is given in Section \ref{sec_data}. 
Section \ref{sec_system} describes the \gls{DPCL} and \gls{ASR} system, before the experimental setup and results are presented in Section \ref{sec_experiments} and discussed in Section \ref{sec_discussion}. 
Potential future research directions are discussed in Section \ref{sec_conclusion}.

\section{Data}
\label{sec_data}
We report \gls{ASR} results on the commonly used dataset \mbox{wsj0-2mix} introduced in \cite{deeCluDisEmbForSegAndSep}.
This dataset is created by artificial mixing of speakers from the \gls{WSJ} data.
The main drawback of this datasets for \gls{ASR} experiments is, that all generated utterances contain fully overlapping speech.
This means that both speakers talk for almost the complete length of the utterance.
In a realistic \gls{ASR} scenario for overlapping speech, as they can be found e.g. in the CHiME-5 or AMI datasets \cite{theFifChiSpeSepAndRecChaDatTasAndBas, theAmiMeeCorAPreAnn}, speakers are only overlapping for smaller portions of an utterance.
This means that each utterance contains significant portions of single-speaker speech.

To study this scenario of more sparsely overlapping speech in the effect of \gls{DPCL} as a preprocessing step for \gls{ASR}, we create datasets containing sparsely overlapping speech utterances. 
Other aspects of the artificial mixing, such as the \gls{SNR} distribution, are kept as similar as possible to the algorithm used in \cite{deeCluDisEmbForSegAndSep}.
The data simulation pipeline is described in the following.

Two separate signal tracks are generated, each containing speech from a single speaker.
The \gls{ASR} system described in Section \ref{sec_systemAsr} is used to obtain a forced alignment for the source datasets, where the speech segments are sampled from.
This alignment is used to cut leading and trailing silences from the utterances.
This ensures that only pauses in between words remains as silence in those utterances, which is neglectable in the computation of the overlap ratio.
Furthermore, a silence set containing those leading and trailing silence segments is created.

For each speaker three utterances are sampled from the source dataset.
One signal track is created for each of the two speakers where the sampled segments are separated by silence gaps.
The lengths of the silence gaps are randomly sampled with the constraint that the overlap of speech after adding the two signal tracks has a given overlap ratio and that the ratio of the mixed signal containing no speech does not exceed a certain threshold (here \SI{10}{\%}).
The silence gaps are then filled with silence signals sampled from the silence set mentioned above, where the energy of the silence segments used to fill the gap is scaled to the leading and trailing silences of the original speech segments.
The two signal tracks are then mixed with a given \gls{SNR} value similar to the data simulation of \mbox{wsj0-2mix}, where the signal energy of the two signal tracks is computed by only considering the speech segments and not the silence segments.

\section{System}
\label{sec_system}
\subsection{ASR system}
\label{sec_systemAsr}

A state of the art hybrid \gls{DNN}-\gls{HMM} acoustic model, trained on the WSJ-SI84 subset (\SI{15}{h}) of the \gls{WSJ} dataset, is used for the experiments. 
The input features are unnormalized 80 dimensional log-Mel filterbank features based on a \gls{STFT} employing the Hanning window applied to a \SI{25}{ms} frame with a frame shift of \SI{10}{ms}.
Since the input features are unnormalized the first layer of the acoustic model is an 80 dimensional linear layer employing batch normalization \cite{batNorAccDeeNetTraByRedIntCovShi}.
The linear layer is followed by 5 \gls{BLSTM} layers with 600 units each.
The output is a softmax layer with 1501 units.
A 3-gram language model is used during recognition.
Table \ref{tbl_origWsj} shows the \gls{ASR} performance of the system on the standard 5k and 20k development and evaluation datasets of \gls{WSJ}, using the respective language model is used.

The system is implemented using RETURNN and RASR \cite{retTheRwtExtTraFraForUniRecNeuNet, rasNnTheRwtNeuNetTooForSpeRec}.

\begin{table}[!h] 
\centering 
\caption{WER in \% of the ASR system on the standard development and evaluation sets of the WSJ dataset. The data contained in the sets is different for 5k and 20k.} 
\label{tbl_origWsj} 
\setlength\tabcolsep{4pt} 
\begin{tabular}{|c|c|c|c|}     
\hline     
\thead{Vocabulary \\ size} & \thead{dev93} & \thead{eval92} & \thead{eval93} \\     
\hline
5k & \wsjOrigDevNinetythreeFiveK & \wsjOrigEvalNinetytwoFiveK & \wsjOrigEvalNinetythreeFiveK \\
\hline     
20k & \wsjOrigDevNinetythreeTwentyK & \wsjOrigEvalNinetytwoTwentyK & \wsjOrigEvalNinetythreeTwentyK \\
\hline     
\end{tabular}
\end{table}

\subsection{Source separation system}
\label{sec_sourceSeparationSystem}

Source separation and \gls{ASR} are handled in two separate stages.
The source separation is done by applying \gls{DPCL} to the mixed speech signal creating one signal per speaker, which is referred to as speaker track in the following. 
The speaker tracks are then fed separately into the \gls{ASR} system.
The following sections give a quick summary of \gls{DPCL} and how it is applied here. 

\subsubsection{Deep clustering network}
The network architecture for \gls{DPCL} described in \cite{sinChaMulSpeSepUsiDeeClu} was reimplemented using RETURNN \cite{retTheRwtExtTraFraForUniRecNeuNet}.
The architecture consists of an \gls{ANN}, which computes a 40 dimensional embedding vector for each time-frequency bin of the input signal.
As input features the \gls{STFT} of the input signal is computed with a window size of \SI{32}{ms}, a frame shift of \SI{8}{ms} and a \gls{DFT} of dimension 512 is used as input features.
The embedding vectors are used to cluster the time-frequency bins into multiple classes (one for each speaker) using soft clustering.
A binary mask is generated from the classification of the time-frequency bins. 
Those masks are applied to the input signal obtaining a separate speaker track for each speaker.
The resulting signals are the input to an enhancement network as described in \cite{sinChaMulSpeSepUsiDeeClu}.
The architecture of the embedding network consists of 4 \gls{BLSTM} layers with 600 units each. 
Curriculum learning is applied as described in \cite{sinChaMulSpeSepUsiDeeClu} with an input size of 100 frames for 100 epochs and 400 frames for 100 epochs.
The architecture of the enhancement network consists of 2 \gls{BLSTM} layers with 300 units each. 
The \gls{SDR} improvements obtained by this systems on the \mbox{wsj0-2mix} dataset are shown in Table \ref{tbl_wsj02mixOriginalGenderEval} and are in line with the results described in \cite{sinChaMulSpeSepUsiDeeClu}.

\subsubsection{Application of DPCL to sparsely overlapping speech}
\label{sec_dpclOnSparselyOverlappingData}
Applying \gls{DPCL} to sparsely overlapping speech signals as described in Section \ref{sec_data} can be done in the same manner as for the fully overlapping speech.
This approach is referred to as full-sequence mode hereafter.
This approach can potentially suffer from signal quality degradation of the single speech segments, due to erroneous masking.
An alternative approach is to apply \gls{DPCL} to the multi-speaker segments only.
This second approach requires to deal with the segmentation problem, meaning how to separate the input signal into multi-speaker and single-speaker segments.
We experimented with various ways to solve the segmentation problem, but those results will be presented in future work, since they go beyond the scope of this work and do not serve to further the conclusions presented here.
The results reported here use the oracle knowledge for segmentation.

The embedding vectors are computed for the complete signal.
The single-speaker segments remain unprocessed, while one output signal per speaker is generated for the multi-speaker segments by computing the masks based on the embeddings of only that segment.
This creates a segment permutation problem, where the resulting output segments need to be allocated to an output speaker track.
For the experiments presented here a fixed number of 2 speaker tracks is used.

Three approaches to handle the permutation problem are used in this work.
First the oracle knowledge is used.
This is done by computing the correlations to the respective segment of the source signals tracks for each of the outputs per segment.
The output segment is allocated to the source signal track with the higher correlation and thus to a speaker track.

The second approach is hereafter referred to as affinity approach.
In this approach the mean of the embedding vectors for each speaker in the multi-speaker segments is calculated.
For each possible permutation of multi-speaker segments the average distance of the resulting group of mean vectors is computed and the permutation with the lowest average distance is selected.
Then a mean embedding vector for each speaker track based on the selected permutation is computed.
The single-speaker segments are then allocated to the speaker track which mean embedding vector is closest to the mean embedding vector of the single-speaker segment. 

The third approach is hereafter referred to as speaker-Id approach.
In this approach the \gls{DPCL} network is trained in a multi-task approach similar to \cite{deeAttNetForSpeReIdeAndBliSouSep}.
The network is extended by a second output which is utilized for speaker identification. 
Different embedding vectors are computed for the speaker identification part of the network.
Those speaker-Id embedding vectors are then used to handle the permutation problem in the same manner as is done in the affinity approach.
Details about the network architecture and the cost function are described in \cite{deeAttNetForSpeReIdeAndBliSouSep}. 
The main difference of our network architecture to the one presented in \cite{deeAttNetForSpeReIdeAndBliSouSep} is that the deep attractor network from \cite{deeAttNetForSpeReIdeAndBliSouSep} is replaced by the \gls{DPCL} network described above and the dimension of the speaker-Id embedding vectors which is 40.
Furthermore both cost functions are weighted equally in the multi-task training.


\section{Experimental setup and results}
\label{sec_experiments}
Table \ref{tbl_wsj02mixOriginal} shows the \gls{WER} of the system on the fully overlapping dataset \mbox{wsj0-2mix} described in Section \ref{sec_data}.
The table also shows the more recently published performance of sequence-to-sequence systems.
In past publications those systems have only been compared to a system employing \gls{DPCL} in combination with a \gls{GMM}-\gls{HMM} acoustic model \cite{sinChaMulSpeSepUsiDeeClu} and have been shown to yield better \glspl{WER} \cite{aPurEndToEndSysForMulSpeSpeRec, endToEndMonMulSpeAsrSysWitPre}.
But the results in Table \ref{tbl_wsj02mixOriginal} show \gls{DPCL} to be superior to the integrated approaches, when combined with a state of the art acoustic model. 


\begin{table}[!h] 
\centering 
\caption{WER in \% of the described system compared to recently published results on wsj0-2mix. The acoustic models from \cite{aPurEndToEndSysForMulSpeSpeRec} and \cite{endToEndMonMulSpeAsrSysWitPre} differ in details.} 
\label{tbl_wsj02mixOriginal} 
\setlength\tabcolsep{4pt} 
\begin{tabular}{|c|c|c|c|c|}     
\hline     
\multicolumn{4}{|c|}{System} & \multirow{2}{*}{Eval} \\     
\cline{1-4}
\thead{Separation} & \thead{\thead{Acoustic\\ model}} &\thead{\thead{Language\\ model}} & \thead{Reference} & \\     
\hline
\multirow{2}{*}{\thead{DPCL}} & \thead{GMM-HMM} & n/a & \cite{sinChaMulSpeSepUsiDeeClu} & \dpclGmmHmmEvalA \\
\cline{2-5}
                      & \thead{DNN-HMM} & \thead{3-gram} & \thead{proposed\\ here} &  \bf \dpclDnnHmmEvalA \\
\hline
\multirow{2}{*}{integrated} & \multirow{2}{*}{\thead{joint\\CTC/\\attention}} & \thead{word- \& \\char-level \\RNNLMs} & \cite{aPurEndToEndSysForMulSpeSpeRec} & \EtEOneEvalA \\
\cline{3-5}
                            & & \thead{word--level \\RNNLMs} & \cite{endToEndMonMulSpeAsrSysWitPre} & \EtETwoEvalA \\
\hline
\end{tabular}
\end{table}

The results in Table \ref{tbl_wsj02mixOriginalGenderEval} show, that \gls{DPCL} is much more reliable, when the competing speaker is of different gender to the dominant speaker. 
This effect can also be seen in the signal quality metric \gls{SDR} as presented in previous publications \cite{sinChaMulSpeSepUsiDeeClu} and confirmed by our experiments. 
In our experiments the effect seems to be slightly stronger for \gls{WER} than for \gls{SDR}.

\begin{table}[!h] 
\centering 
\caption{WER and SDR of the proposed system differentiated by gender of competing speaker and speaker dominance.} 
\label{tbl_wsj02mixOriginalGenderEval} 
\setlength\tabcolsep{4pt} 
\begin{tabular}{|c|c|c|c|}     
\hline     
\thead{Speaker} & \thead{Gender} & \thead{Eval WER (\%)} & \thead{Eval SDR} \\     
\hline
\multirow{2}{*}{dominant} & same & \dpclDnnHmmSameDomEvalA & \dpclDnnHmmSameDomEvalASdr \\
\cline{2-4}
 & diff & \dpclDnnHmmDiffDomEvalA & \dpclDnnHmmDiffDomEvalASdr \\
\hline
\multirow{2}{*}{non dominant} & same & \dpclDnnHmmSameNondomEvalA & \dpclDnnHmmSameNondomEvalASdr \\
\cline{2-4}
 & dff & \dpclDnnHmmDiffNondomEvalA & \dpclDnnHmmDiffNondomEvalASdr \\
\hline
\end{tabular}
\end{table}

The drawback of the \mbox{wsj0-2mix} datasets is, that it does not cover scenarios in which a significant portion of the speech signal contains single-speaker segments and only part of the signal contains multi-speaker segments.
This is what one would e.g. expect in meeting recordings or the smart home scenario. 
The following experiments investigate the additional obstacles that those scenarios pose for the application of \gls{DPCL} as a preprocessing for \gls{ASR}.

The data used in the following experiments has been created as described in Section \ref{sec_data}.
The data for the evaluation set is sampled from \mbox{\tt si\_dt\_05} and \mbox{\tt si\_et\_05}, which are both not used for training, cross validation or hyper parameter tuning.
We chose those source datasets to stay as close as possible to the original \mbox{wsj0-2mix} dataset presented in Table \ref{tbl_wsj02mixOriginal} and \ref{tbl_wsj02mixOriginalGenderEval} to make the following results most comparable to the fully overlapping scenarios.
As before decoding was done using a 3-gram language model with a vocabulary size of 20k.

\begin{figure*}[th!]
\centering
\begin{subfigure}{0.32\textwidth}
\includegraphics[width=\textwidth]{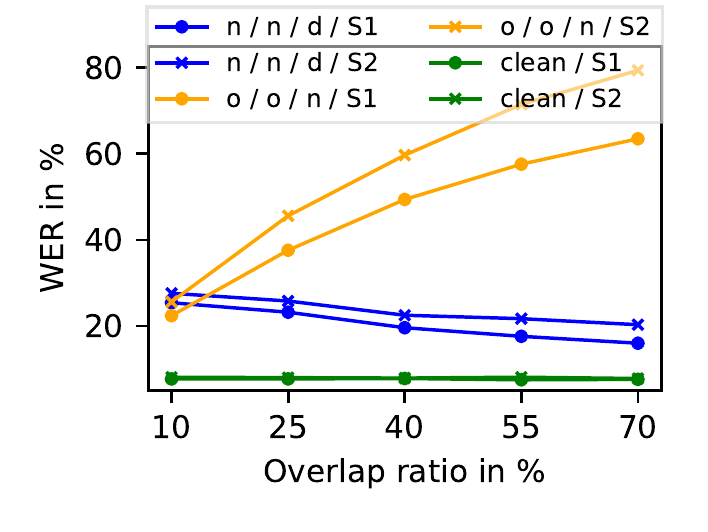}
\vspace{-0.8cm}
\caption{Complete set}
\label{fig_plotsWerOverOr_fullSeqBoth}
\end{subfigure}
\hfil
\begin{subfigure}{0.32\textwidth}
\includegraphics[width=\textwidth]{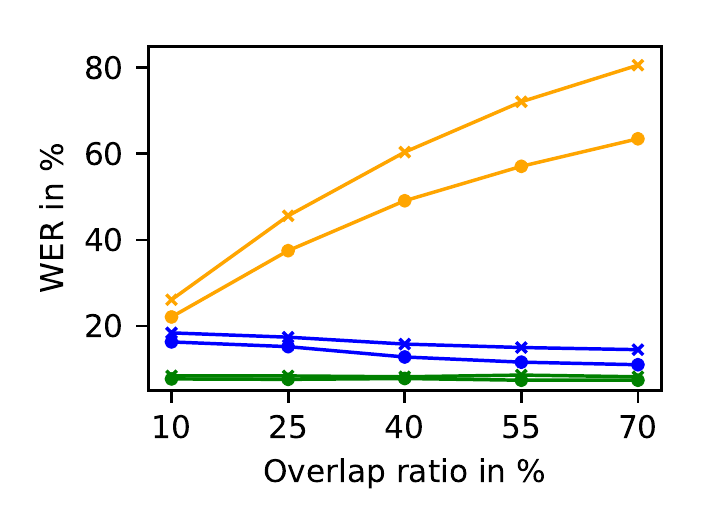}
\vspace{-0.8cm}
\caption{Competing speaker of different gender}
\label{fig_plotsWerOverOr_fullSeqDiff}
\end{subfigure}
\hfil
\begin{subfigure}{0.32\textwidth}
\includegraphics[width=\textwidth]{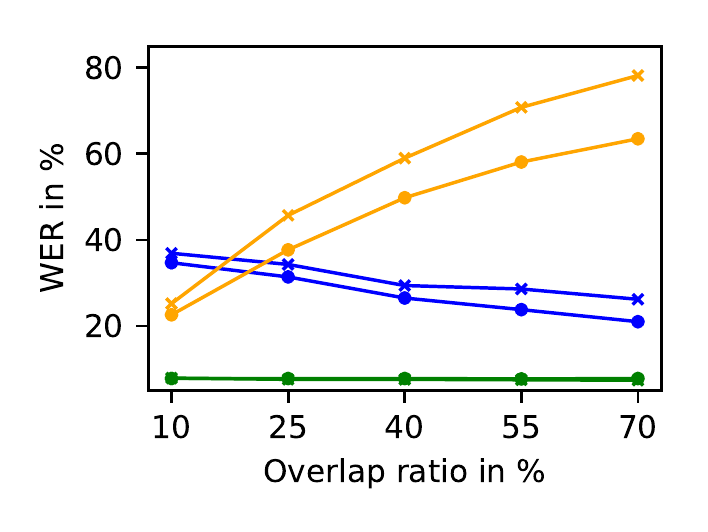}
\vspace{-0.8cm}
\caption{Competing speaker of same gender}
\label{fig_plotsWerOverOr_fullSeqSame}
\end{subfigure}
\begin{subfigure}{0.32\textwidth}
\includegraphics[width=\textwidth]{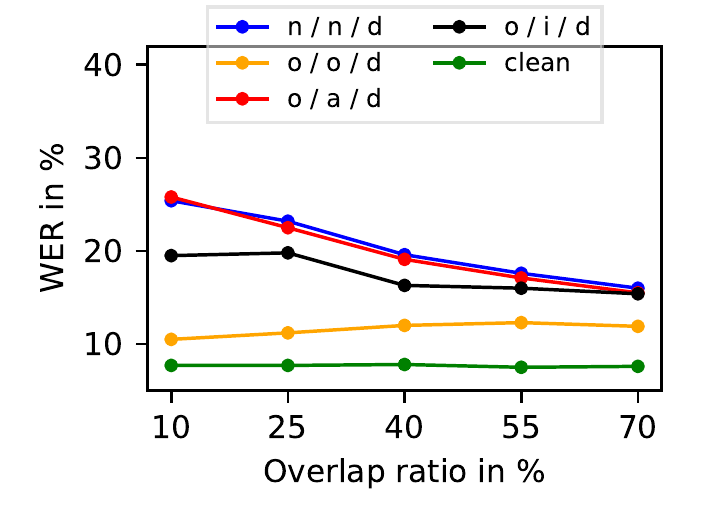}
\vspace{-0.8cm}
\caption{Complete set}
\label{fig_plotsWerOverOr_oracleBoth}
\end{subfigure}
\hfil
\begin{subfigure}{0.32\textwidth}
\includegraphics[width=\textwidth]{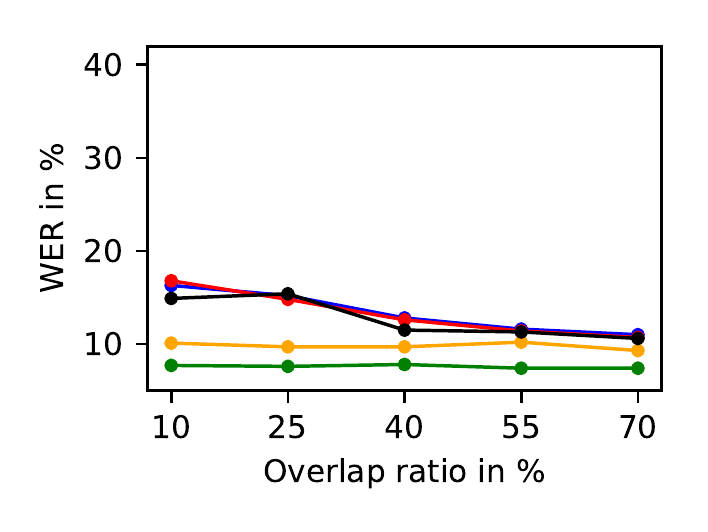}
\vspace{-0.8cm}
\caption{Competing speaker of different gender}
\label{fig_plotsWerOverOr_oracleDiff}
\end{subfigure}
\hfil
\begin{subfigure}{0.32\textwidth}
\includegraphics[width=\textwidth]{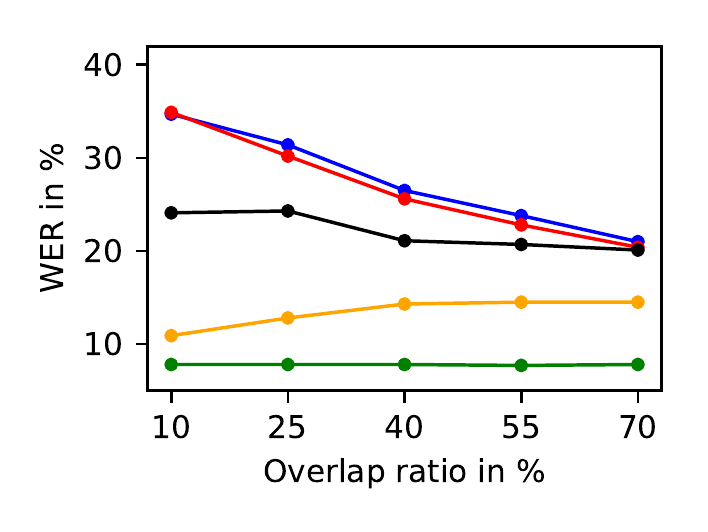}
\vspace{-0.8cm}
\caption{Competing speaker of same gender}
\label{fig_plotsWerOverOr_oracleSame}
\end{subfigure}
\caption{Analysis of DPCL for various overlap ratios. a)-c) compare DPCL to an upper and a lower performance bound. d)-f) compare, for the dominant speaker, the application of DPCL in full-sequence mode and the application of DPCL on the multi-speaker segments only. The legend syntax is as follows: \\
segmentation / permutation / source separation [ / speaker dominance ]\\
n = none,
o = oracle,
a = affinity permutation,
i = speaker-Id permutation,
d = DPCL,
S1 = dominant speaker,
S2 = non dominant speaker
}
\label{fig_plotsWerOverOr}
\end{figure*}

The \gls{WER} on the separate signal tracks before mixing can be considered a lower boundary for the \glspl{WER} and is referred to as clean in the following. 
The \gls{WER} on the mixed signal is dominated by insertion errors induced by the single-speaker speech segments of the competing speaker. 
Therefore this \gls{WER} is less useful to investigate degradation which stem from the multi-speaker segments.
An alternative sensible upper reference for the \gls{WER} can be described by a perfect speaker identification system, which allocates multi-speaker segments to both speakers.
In our experiments this is the same as not applying any source separation to the multi-speaker segments and using oracle knowledge for segmentation and permutation.
Figure \ref{fig_plotsWerOverOr} shows the \glspl{WER} of the various processing approaches over the evaluation sets with various overlap ratios.

\section{Discussion}
\label{sec_discussion}

As expected the \gls{ASR} performance strongly improves with decreasing overlap ratio when no separation is applied to the multi-speaker segments and oracle knowledge for segmentation and permutation is used.
On the other hand the \gls{ASR} performance when applying \gls{DPCL} in full-sequence mode, decreases for decreasing overlap ratios.
The results shows that even though \gls{DPCL} works extremely well on fully overlapping speech a simple direct application to sparsely overlapping speech could potentially even hurt \gls{ASR} performance compared to no processing of the mixed signal.

Figure \ref{fig_plotsWerOverOr_oracleBoth} shows, that if oracle knowledge is used for the segmentation and permutation problem and \gls{DPCL} is only applied to the multi-speaker segments, the \gls{WER} improves for decreasing overlap ratios.
The gap between the application of \gls{DPCL} in full-sequence mode and its application to only the multi-speaker segments with utilization of oracle knowledge for segmentation and permutation shows the maximum potential performance gain that can be obtained if the segmentation and permutation problems are solved optimally.
Especially for low overlap ratios the potential for improving \gls{ASR} performance is large.
Closing this gap is crucial for the applicability of \gls{DPCL} in real world \gls{ASR}.

The intuitive expectation when applying \gls{DPCL} to only the multi-speaker segments with utilization of oracle knowledge for segmentation and permutation is an increase in \gls{WER} for increasing overlap ratios, since larger portions of the signal will suffer from quality degradation due to the required source separation.
But Figure \ref{fig_plotsWerOverOr_oracleBoth}, shows only a minor decrease in \gls{ASR} performance over increasing overlap ratios.
A differentiation of scenarios in which the competing speaker has the same or different gender as the dominant speaker as done in Figures \ref{fig_plotsWerOverOr_oracleDiff} and \ref{fig_plotsWerOverOr_oracleSame} reveal that the decrease in \gls{ASR} performance for larger overlap ratios stems almost exclusively from the same gender scenario.

Furthermore the performance of \gls{DPCL} applied in full-sequence mode is significantly worse for the same gender scenario throughout all overlap ratios.
One explanation for the higher difficulty of the same gender scenario is that mask based separation approaches rely on sparsity of the acoustic features along the frequency domain, which is more dominant in the different gender scenario.
This can explain why, in the same gender scenario, the performance increases more strongly with decreasing overlap ratio, when applying \gls{DPCL} to only the multi-speaker segments with utilization of oracle knowledge for segmentation and permutation .
If the sparsity is also the main obstacle for the application of \gls{DPCL} on the full-sequence mode, the performance gap between the different gender scenario and the same gender scenario should shrink for decreasing overlap ratios.
This is not the case as can be seen in Figures \ref{fig_plotsWerOverOr_oracleDiff} and \ref{fig_plotsWerOverOr_oracleSame}.
This indicates that the main problem is not the masking of the multi-speaker segments, but the handling of single-speaker segments.

When the oracle permutation is replaced by the affinity approach described in Section \ref{sec_sourceSeparationSystem} it can be observed that the \glspl{WER} are very similar to the \glspl{WER} of applying \gls{DPCL} in full-sequence mode.
Combined with the findings described above, this indicates that the main obstacle for the application of \gls{DPCL} to sparsely overlapping speech is not the potential signal degradation from the masking of the single- or multi-speaker segments, but the collective allocation of the time-frequency bins of the single-speaker segments to a speaker in the multi-speaker segments.

A straight forward solution to this problem could be to provide the \gls{DPCL} network with sparsely overlapping data during training.
To this end we trained multiple \gls{DPCL} networks with training data containing various average overlap ratios.
The data was simulated as described in Section \ref{sec_data} and the network was trained as described in Section \ref{sec_sourceSeparationSystem}.
The total amount of source data used for the simulation has been kept constant for a fair comparison with the network trained on fully overlapping signals.
Without adjustment of the training cost function we were not able to gain an improvement by only changing the training data in that manner.
Investigations into this approach will be future work.

A different solution is to improve the handling of the permutation problem when handling segmentation, permutation and source separation separately. 
For this we utilize the speaker-Id approach described in Section \ref{sec_dpclOnSparselyOverlappingData}.
With this approach we were able to get improvements of about \SI{20}{\%} relative for low overlap ratios.
It can also be seen that this improvement is mainly due to the same gender scenario and that the effect vanishes for higher overlap ratios.

\section{Conclusions}
\label{sec_conclusion}

In this work we have shown that the use of \gls{DPCL} as a separate speaker separation step for multi-speaker \gls{ASR} works well on the standard \mbox{wsj0-2mix} dataset if it is combined with a state of the art \gls{DNN}-\gls{HMM} acoustic model.
To the best of our knowledge the \gls{WER} of \SI{\dpclDnnHmmEvalA}{\%} obtained by this system is currently the lowest \gls{WER} reported on wsj0-2mix.
Furthermore we presented in depth investigations on the effects of \gls{DPCL} as a preprocessing step for \gls{ASR} on sparsely overlapping speech. 
To this end we simulated data with varying overlap ratio of the competing speakers.
Those experiments on simulated data aim to further the utilization of \gls{DPCL} to improve \gls{ASR} performance on real data as e.g. the AMI or CHiME-5 datasets.

The results presented here indicate the main obstacles to obtain similar \gls{ASR} gains on real data as can be seen for the \mbox{wsj0-2mix} data.
More specifically it has been shown that a major drawback of the basic \gls{DPCL} approach is the handling of single-speaker segments.
The results indicate that the main reason is not a degradation of signal quality of the single-speaker segments by erroneous masking,
but rather the problem of allocating the time-frequency bins of a single-speaker segment to a speaker of the multi-speaker segments.
The results show that a promising approach is to separate the source separation into three different steps.
The first step is the segmentation of the signal into single-speaker and multi-speaker segments.
In a second step \gls{DPCL} is applied to the multi-speaker segments only.
Finally the problem of allocating the resulting segments to a speaker track needs to be solved. 
For this problem we have presented an approach based on speaker identification, which improved the results more than \SI{20}{\%} relative for low overlap ratios. 

Future work will focus on the modification of \gls{DPCL} to obtain a better handling of single-speaker segments in the full-sequence approach. 
This could be attempted by introducing a regularizing term in the cost function during training on sparsely overlapping training data.
Furthermore we will explore the use of feedback from the acoustic model to solve the segmentation and permutation problems in the separated approach.

\vspace{0.9\baselineskip}
\begin{acknowledgements}
This project has received funding from the European Research Council (ERC) under the European Union's Horizon 2020 research and innovation program grant agreement No. 694537 and under the Marie Sk\l{}odowska-Curie grant agreement No. 644283 and from a Google Focused Award.
The work reflects only the authors' views and none of the funding agencies is responsible for any use that may be made of the information it contains.
\end{acknowledgements}

\newpage

\bibliographystyle{IEEEtran}

\linespread{1.0}

\bibliography{mybib}

\begin{thebibliography}{10}
\providecommand{\url}[1]{#1}
\csname url@samestyle\endcsname
\providecommand{\newblock}{\relax}
\providecommand{\bibinfo}[2]{#2}
\providecommand{\BIBentrySTDinterwordspacing}{\spaceskip=0pt\relax}
\providecommand{\BIBentryALTinterwordstretchfactor}{4}
\providecommand{\BIBentryALTinterwordspacing}{\spaceskip=\fontdimen2\font plus
\BIBentryALTinterwordstretchfactor\fontdimen3\font minus
  \fontdimen4\font\relax}
\providecommand{\BIBforeignlanguage}[2]{{%
\expandafter\ifx\csname l@#1\endcsname\relax
\typeout{** WARNING: IEEEtran.bst: No hyphenation pattern has been}%
\typeout{** loaded for the language `#1'. Using the pattern for}%
\typeout{** the default language instead.}%
\else
\language=\csname l@#1\endcsname
\fi
#2}}
\providecommand{\BIBdecl}{\relax}
\BIBdecl

\bibitem{theMic2017ConSpeRecSys}
W.~Xiong, L.~Wu, F.~Alleva, J.~Droppo, X.~Huang, and A.~Stolcke, ``The
  microsoft 2017 conversational speech recognition system,'' in \emph{Proc.
  IEEE International Conference on Acoustics, Speech and Signal Processing
  (ICASSP)}, Calgary, Canada, Apr 2018, pp. 5934--5938.

\bibitem{theCap2017ConvSpeRecSys}
K.~J. Han, A.~Chandrashekaran, J.~Kim, and I.~Lane, ``The {CAPIO} 2017
  conversational speech recognition system,'' \emph{arXiv preprint
  arXiv:1801.00059}, 2017.

\bibitem{theFifChiSpeSepAndRecChaDatTasAndBas}
J.~Barker, S.~Watanabe, E.~Vincent, and J.~Trmal, ``The fifth {'CHiME'} speech
  separation and recognition challenge: Dataset, task and baselines,'' in
  \emph{Proc. Interspeech}, Hyderabad, India, Sep 2018, pp. 1561--1565.

\bibitem{theAmiMeeCorAPreAnn}
J.~Carletta, S.~Ashby, S.~Bourban, M.~Flynn, M.~Guillemot, T.~Hain, J.~Kadlec,
  V.~Karaiskos, W.~Kraaij, M.~Kronenthal, B.~Lathoud, M.~Lincoln, A.~Lisowska,
  I.~McCowan, W.~Post, D.~Reidsma, and P.~Wellner, ``The {AMI} meeting corpus:
  A pre-announcement,'' in \emph{International Workshop on Machine Learning for
  Multimodal Interaction}.\hskip 1em plus 0.5em minus 0.4em\relax Springer,
  2005, pp. 28--39.

\bibitem{perInvTraOfDeeModForSpeIndMulTalSpeSep}
D.~Yu, M.~Kolb{\ae}k, Z.-H. Tan, and J.~Jensen, ``Permutation invariant
  training of deep models for speaker-independent multi-talker speech
  separation,'' in \emph{Proc. IEEE International Conference on Acoustics,
  Speech and Signal Processing (ICASSP)}, New Orleans, LA, USA, Mar 2017, pp.
  241--245.

\bibitem{aPurEndToEndSysForMulSpeSpeRec}
H.~Seki, T.~Hori, S.~Watanabe, J.~Le~Roux, and J.~R. Hershey, ``A purely
  end-to-end system for multi-speaker speech recognition,'' in \emph{Proc. of
  the 56th Annual Meeting of the Association for Computational Linguistics
  (ACL)}, Melbourne, Australia, 2018, pp. 2620--2630.

\bibitem{endToEndMonMulSpeAsrSysWitPre}
X.~Chang, Y.~Qian, K.~Yu, and S.~Watanabe, ``End-to-end monaural multi-speaker
  {ASR} system without pretraining,'' in \emph{accpeted for publication in
  Proc. IEEE International Conference on Acoustics, Speech and Signal
  Processing (ICASSP)}, Brighton, UK, May 2019.

\bibitem{deeCluDisEmbForSegAndSep}
J.~Hershey, Z.~Chen, J.~Le~Roux, and S.~Watanabe, ``Deep clustering:
  Discriminative embeddings for segmentation and separation,'' in \emph{Proc.
  IEEE International Conference on Acoustics, Speech and Signal Processing
  (ICASSP)}, Shanghai, China, Mar 2016, pp. 31--35.

\bibitem{mulTalSpeSepWitUttLevPerInvTraOfDeeRecNeuNet}
M.~Kolb{\ae}k, D.~Yu, Z.-H. Tan, and J.~Jensen, ``Multitalker speech separation
  with utterance-level permutation invariant training of deep recurrent neural
  networks,'' \emph{IEEE/ACM Transactions on Audio, Speech, and Language
  Processing}, vol.~25, no.~10, pp. 1901--1913, Oct 2017.

\bibitem{recMulTalSpeWitPerInvTra}
D.~Yu, X.~Chang, and Y.~Wian, ``Recognizing multi-talker speech with
  permutation invariant training,'' in \emph{Proc. Interspeech}, Stockholm,
  Sweden, Aug 2017, pp. 2456--2460.

\bibitem{sinChaMulSpeSepUsiDeeClu}
Y.~Isik, J.~L. Roux, Z.~Chen, S.~Watanabe, and J.~R. Hershey, ``Single-channel
  multi-speaker separation using deep clustering,'' in \emph{Proc.
  Interspeech}, San Francisco, CA, USA, 2016, pp. 545--549.

\bibitem{deeAttNetForSinMicSpeSep}
Z.~Chen, Y.~Luo, and N.~Mesgarani, ``Deep attractor network for
  single-microphone speaker separation,'' in \emph{Proc. IEEE International
  Conference on Acoustics, Speech and Signal Processing (ICASSP)}, New Orleans,
  LA, USA, Mar 2017, pp. 246--250.

\bibitem{endToEndMulSpeSpeRec}
S.~Settle, J.~Le~Roux, T.~Hori, S.~Watanabe, and J.~R. Hershey, ``End-to-end
  multi-speaker speech recognition,'' in \emph{Proc. IEEE International
  Conference on Acoustics, Speech and Signal Processing (ICASSP)}, Calgary,
  Canada, 2018, pp. 4819--4823.

\bibitem{batNorAccDeeNetTraByRedIntCovShi}
S.~Ioffe and C.~Szegedy, ``Batch normalization: Accelerating deep network
  training by reducing internal covariate shift,'' \emph{arXiv preprint
  arXiv:1502.03167}, Mar 2015.

\bibitem{retTheRwtExtTraFraForUniRecNeuNet}
P.~Doetsch, A.~Zeyer, P.~Voigtlaender, I.~Kulikov, R.~Schl{\"u}ter, and H.~Ney,
  ``{RETURNN}: The {RWTH} extensible training framework for universal recurrent
  neural networks,'' in \emph{Proc. IEEE International Conference on Acoustics,
  Speech and Signal Processing (ICASSP)}, New Orleans, LA, USA, Mar 2017, pp.
  5345--5349.

\bibitem{rasNnTheRwtNeuNetTooForSpeRec}
S.~Wiesler, A.~Richard, P.~Golik, R.~Schl{\"u}ter, and H.~Ney, ``{RASR/NN}: The
  {RWTH} neural network toolkit for speech recognition,'' in \emph{Proc. IEEE
  International Conference on Acoustics, Speech and Signal Processing
  (ICASSP)}, Florence, Italy, May 2014, pp. 3281--3285.

\bibitem{deeAttNetForSpeReIdeAndBliSouSep}
L.~Drude, T.~von Neumann, and R.~Haeb-Umbach, ``Deep attractor networks for
  speaker re-identification and blind source separation,'' in \emph{Proc. IEEE
  International Conference on Acoustics, Speech and Signal Processing
  (ICASSP)}, Calgary, Canada, Apr 2018, pp. 11--15.

\end{thebibliography}

\end{document}